\documentstyle{article}

\begin{document}
\input{psfig}

\title{Conformation changes and folding of proteins mediated 
                                        by Davidov's soliton}
\author{Shay Caspi and Eshel Ben-Jacob \\
        School of Physics and Astronomy, Faculty of Exact Sciences,\\ 
        Tel Aviv University, 69978 Tel Aviv, Israel.}

\maketitle

\baselineskip 18pt

\begin{abstract}
We suggest that Davidov's solitons, propagating through the backbone of a protein, can mediate
conformational transition and folding of a protein to its native state. A simple toy model is 
presented in which a Non Linear Schr\"{o}dinger (NLS) field interacts with another field $\phi$ 
corresponding to the conformation angles of the protein. The interaction provides the 
conformation field with the energy needed in order to overcome energy barriers for folding,
thus avoiding the need for stochastic thermal activation. Such a transition is therefore a 
deterministic controlled process. The soliton is compensated for its energy loss by absorption 
of the energy gained in the folding process. This scenario does not change
significantly even in the presence of imposed disorder, provided that the initial momentum of the
soliton is large enough.
\end{abstract}
\vspace*{.2in}

\section{Introduction}
Many properties of biomolecules and their biological functioning can not be fully understood
when only their chemical structure is taken into considerations. Nowadays, it is clear that
biological activity has important physical and dynamical aspects. One concept which has
already made a significant (although somewhat controversial) contribution to the physics
of biological systems is that of solitons. The idea that solitons may provide the mechanism
for energy and charge transport in proteins was developed by Davidov (1977, 1982). 
Since then, many papers that followed Davidov's approach were published 
(for example: Olsen et al. 1989; Ciblis \& Cosic 1997; Cruzeiro-Hansson \& Takeno 1997; 
F\"{o}rner 1997).
In our recent paper we suggested that the folding of proteins to their native state and
the conformational changes in folded proteins can both be associated with the propagation
of solitons through the backbone of the protein (Caspi \& Ben-Jacob 1998). Here we demonstrate 
that such a process can be mediated by explicit, Davidov's-like solitons. 

A protein is a polymer built from 20 different amino acids ordered in sequence. This 
sequence is called the primary structure of that 
protein. The protein's actual three dimensional structure is, however, much more complex. Under
normal conditions it folds to a predetermined and rather static structure. This so called
native state can be described in terms of  
local secondary structure elements such as $\alpha$-helices and $\beta$-sheets, arranged in
a global tertiary structure. Although the secondary and the tertiary structures are 
dictated by the amino acid sequence, it is still not known today how to predict the 3-D structure
from a known sequence (Friesner \& Gunn 1996). 
A possible approach to this problem could be to try and follow the dynamics
of the folding process. Full molecular dynamics simulation of a detailed protein can not exceed,
using existing computation hardware, the micro-second time scale, while the folding process may
take up to minutes, and not less then milliseconds. Moreover, even the
most elaborated simulations can not account for every detail of the
actual physical molecule (they usually neglect quantum effects, for
example). We can never escape the need to determine which properties
of the system are essential and which have no significant
influence. Therefore, simplified models were used in many cases 
(e.g. Camacho \& Thirumalai 1993; Socci \& Onuchic 1994; Shakhnovich 1994; Bruscolini 1997). 

The folding dynamics may be viewed as a navigation process in the energy landscape of the protein 
conformational space. It is believed that the native structure corresponds to a small region in 
that space located in the vicinity of the ground state. As pointed out by Levinthal, a stochastic 
search for the ground state over a random landscape might take cosmological time. From this fact 
one may draw the conclusion that the landscape of natural proteins is not random, but rather 
directed towards the ground state (Wolynes et al. 1995). Overcoming the
energetic barriers is usually ascribed to a stochastic and uncorrelated thermal activation 
(Karplus \& Shakhnovich 1992). We suggested (Caspi \& Ben-Jacob 1998) that other processes, which are
correlated and deterministic, may take dominant roles 
in dictating the folding pathway. These processes involve the interaction between solitonic 
excitations and the conformational degrees of freedom of the protein. In our paper we presented
a simple toy model in which we used a Sine-Gordon topological soliton. Here we shall demonstrate
that folding can also be mediated using a more realistic non-topological Davidov's soliton.

Even folded proteins do not always have a single static conformation. In fact, many
biological processes in living organisms are associated with conformational
changes. The most intensively studied examples are those of heme proteins
such as myoglobin, which have different conformational states depending on whether they are
bound or not to a CO or O$_2$ ligand (Frauenfelder 1988). Hemoglobin, for example has four
subunits, each containing a heme group which is capable of absorbing an O$_2$ molecule. 
Actually, the oxygen binds cooperatively to the hemoglobin (Stryer 1995). 
The affinity for oxygen grows when there are already oxygen molecules bound. The thing that 
causes this change in the affinity is a global transition between two conformational states of the 
hemoglobin, which is induced by the bound oxygen. The exact details of this transition process 
are still not clear, but it is assumed that the bound oxygen and some external conditions (such
as the pH, partial pressure of O$_2$ and CO$_2$, etc.) cause a different conformational state to 
be energetically preferable. Transition from a metastable to a 
stable conformational state is usually treated as being thermally induced. We suggest that 
conformational transition of this sort may be mediated by solitons, via a mechanism
similar to the one we describe for protein folding. We 
shall refer to the process in which conformational changes are induced by the propagation of a 
solitonic wave along the molecular chain as a Soliton Mediated Conformational Transition (SMCT).

Solitons are localized, non-dispersive excitations, which exist in many non-linear systems. They
are very stable, and therefore can propagate without much energy loss or dispersion to
much larger distances than wave-packets of linear waves. 
The role of solitons in the process of folding (or 
change of conformation) in proteins would be to provide an efficient mechanism for extracting
the energy gained in a single event of local conformational transition and transferring 
it to a distant location. There, it may be used to activate another transition, and the 
energy released in that process could be extracted again. This picture is very different from
the usual assumption that the energy released in each folding step dissipates to the environment.  
In a SMCT process, large sections of a protein can fold very fast (actually, instantaneously on a
time scale of the entire global transition). Moreover, the folding process can be orchestrated 
deterministically. By this we do not mean that proteins have specific
initial states. Indeed, the denatured state is an ensemble of many different
conformational states. However, just has we can build an ordered
brick wall from a an unordered pile of bricks by methodically placing
the bricks at their designed places, so can an unordered conformation
be transformed through a well defined path to an ordered one. We shall
demonstrate this point by considering a random initial state in our model.
 
In proteins, the dipole-dipole interaction between neighbouring
amide-I (the CO double bond) quantum modes of vibration gives rise to linear collective
modes known as {\em exitons}. The dipole-dipole interaction is influenced, however, by lattice
vibrations, so the exitons interact with acoustical phonones in the protein. This interaction
introduces non-linearity into the system. Davidov used a variational approach and obtained equations
for the local amplitudes of the quantum modes. In the continuous limit, the equations (after some
transformations) can be approximated by the Non-linear Schr\"{o}dinger equation (NLS):
\begin{equation}
  i\dot{a} + a_{xx} + 2|a|a = 0,
\end{equation}
where the variables were rescaled to non-dimensional units. There has been an intensive study of
this equation in optics, where it is used to describe the propagation of a wave packet envelope in
an optical fiber with a non-linear refractive index. It should be noted, however, that the ``time''
variable in the NLS equation for optical fibers is actually a large-scale spatial variable. In
Davidov's model, $t$ is strictly the time. The NLS equation has solitary wave
solutions that look like wave-packets with a sech-shaped envelop. These are, in fact, 
non-topological solitons, which means that they have no topological constraint. Unlike the
topological Sine-Gordon solitons, which have a fixed amplitude and finite rest-mass (the minimal 
energy needed to create a soliton), the NLS solitons have a velocity-dependent amplitude and
can exist at any arbitrarily small energy.

Although the actual equations that describe the energy and charge transport in proteins are
probably much more complex than the simple NLS equation, they all have similar properties so we 
shall use this equation to model the possible solitonic excitations. Davidov used a similar
formulation to describe solitonic excitation along the hydrogen bonds in an $\alpha$-helix
structure, and this is the model that is most frequently cited. We, however, consider here 
solitons that move along the backbone of the protein. As for Davidov's model of charge
conduction in proteins, we believe that topological solitons of the Sine-Gordon type,
analogous to those in single-strand DNA (Hermon et al. 1998), are more appropriate in this case.

\section{The model}
Following our previous paper (Caspi \& Ben-Jacob 1998), we shall introduce a toy-model inspired by 
the generic 
properties of protein molecules. Proteins have two local angles per residue (usually denoted 
$\phi$ and $\psi$), which are relatively free, and may be considered at this stage as the most
important degrees of freedom. For almost all amino acid residues in a polypeptide chain, there are 
two distinct minima of the local potential energy in the $(\phi, \psi)$ plane, corresponding to 
the $\alpha$ and the $\beta$ local conformation of the chain. The only exceptions
are glycin, which has four minima, and proline which has only one. We shall
use a scalar variable $\phi$ to represent the local conformation of the protein. In the continuous
limit, which we shall employ for simplicity, it corresponds to the local curvature of the 
molecule. The local potential energy will be simply modeled by an asymmetric $\phi^4$
double well potential, namely
\begin{equation}
  V(\phi) = \varepsilon (\phi + \delta)^2(\phi^2 - {2 \over 3}\phi\delta +
  {1 \over 3}\phi^2 - 2),
\end{equation}
where $\delta$ is the asymmetry parameter, ranging from $-1$ to $1$. The two
minima are positioned at $\phi  = \pm 1$. The energy difference between
the minima is $\Delta E = {16 \over 3} \varepsilon \delta$.
The maxima is positioned at $\phi = -\delta$ and its energy is always zero.

We shall now assume that the amide-I vibrations could be influenced by the local conformation
of the protein, and therefore the amplitude field $a(x)$ interacts with the conformation field 
$\phi$. In order that solitons propagating through the protein backbone could mediate changes in 
its conformation, an interaction of an appropriate form should be introduced. Consider the 
following interaction potential:
\begin{equation}
  u(a,\phi) = \Lambda |a|^2\phi^2,
\end{equation}
where $\Lambda$ is a positive parameter. If $a$ is non-zero, then, at the minima, the energy
of the combined $V$ plus $u$ potential increases, which effectively lowers the
barrier for a folding transition. We therefore expect that a sufficiently energetic soliton may 
enable transition from a metastable to a stable conformation.

The full Lagrangian density then reads:
\begin{equation}
\label{eq_Lagrangian}
  {\cal L} = ia^*\dot{a} - |a_x|^2 + |a|^4 +
             {1\over 2}m\dot{\phi}^2 - V(\phi) - u(a,\phi).
\end{equation}
It should be emphasized that this Lagrangian describes only a ``toy protein''. We would like
to get a qualitative understanding of SMCT processes, rather than give a precise but intractable
description of complicated interactions in a realistic model. We ignore global interactions at
this initial stage. Those interactions can be incorporated later by letting the parameters of the
local potential be dependent on the global conformation. Since we expect that the time scale of
global conformational changes would be much longer than of the local SMCT events, the parameters 
should vary relatively slowly in time. If fact, we imagine that a fast
local SMCT event is followed by a relaxation process in which the strains caused by the the
local conformational transition are relaxed by a slow global
transition. In our model we consider only the fast SMCT events in
which the global conformation is almost constant. Our adiabatic approach is therefore justified.  

From Lagrangian (\ref{eq_Lagrangian}) we obtain the following equations of motion
\begin{eqnarray}
\label{eq_of_motion}
  i\dot{a} & = & -a_{xx} - 2|a|^2a + \Lambda\phi^2a \\
  m\ddot{\phi} & = & -4\varepsilon(\phi+\delta)(\phi^2-1) -
           2\Lambda|a|^2\phi - \Gamma\dot{\phi},
\end{eqnarray}
that also include a dissipation term. These equations can be solved numerically. We considered 
initial conditions of a moving soliton, with $\phi=1$ (local minima) for all x. Note that, as
in our SG model (Caspi \& Ben-Jacob 1998), the form of the interaction potential lowers the 
barrier between the 
two local minima of the conformation energy, so it enables transformation between the two states.
Moreover, it is by no means obvious that enough energy, which is released during this transition,
would be returned to the soliton. In order that the soliton could propagate further and the 
folding process could go on, the gained energy should balance the energy extracted 
by the conformation field as it overcomes the energy barrier and the energy lost to 
radiation of non-solitonic NLS modes induced by the interaction. 
Unlike the SG model in which the 
topological constraint ensured the stability of the soliton, here the soliton may be destroyed 
completely when it loses its energy.

\section{Numerical results}
\begin{figure}
  \centerline{\hbox{
    \psfig{figure=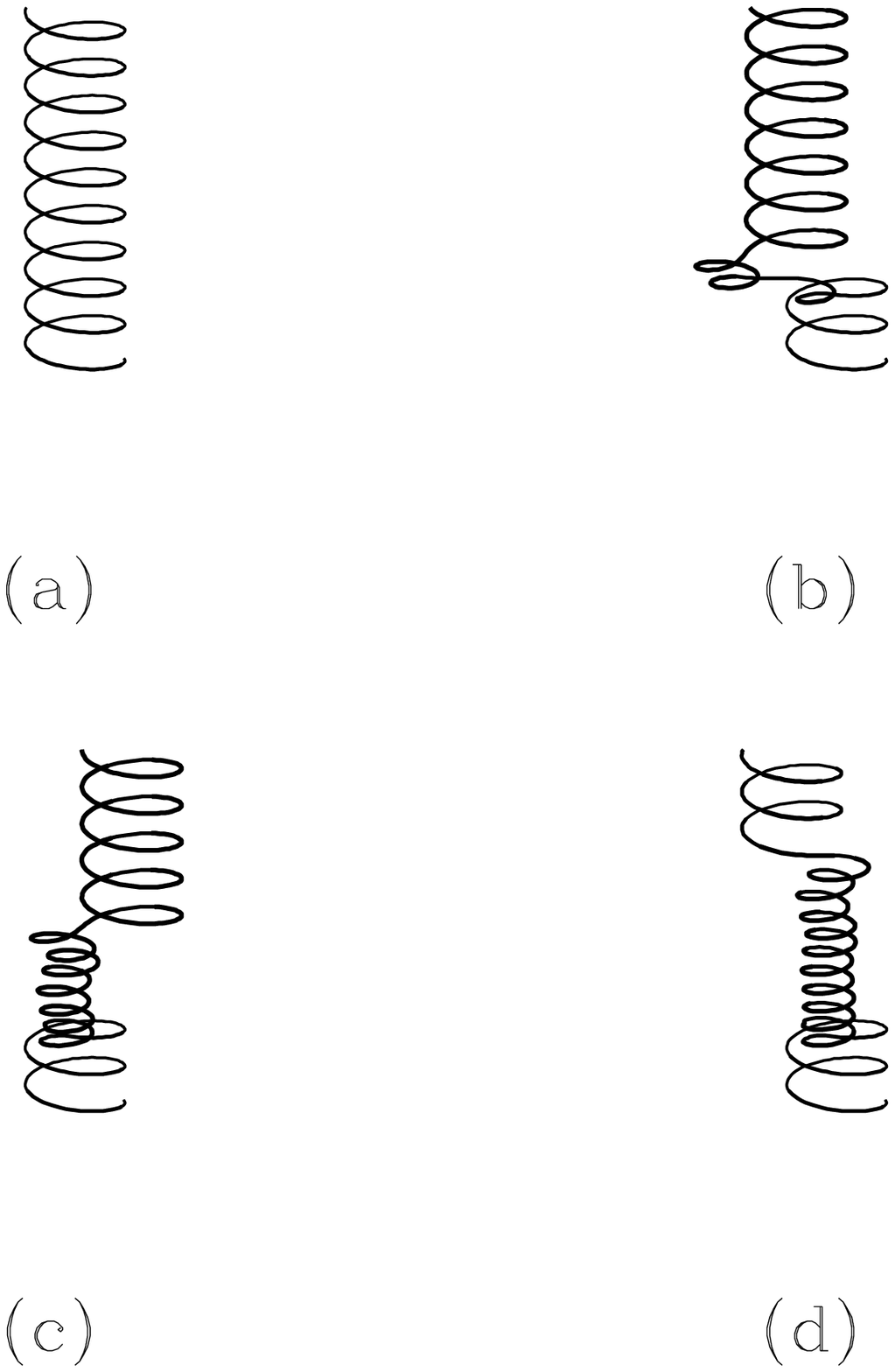,height=6.5cm}
    \hspace{0.2in}
    \psfig{figure=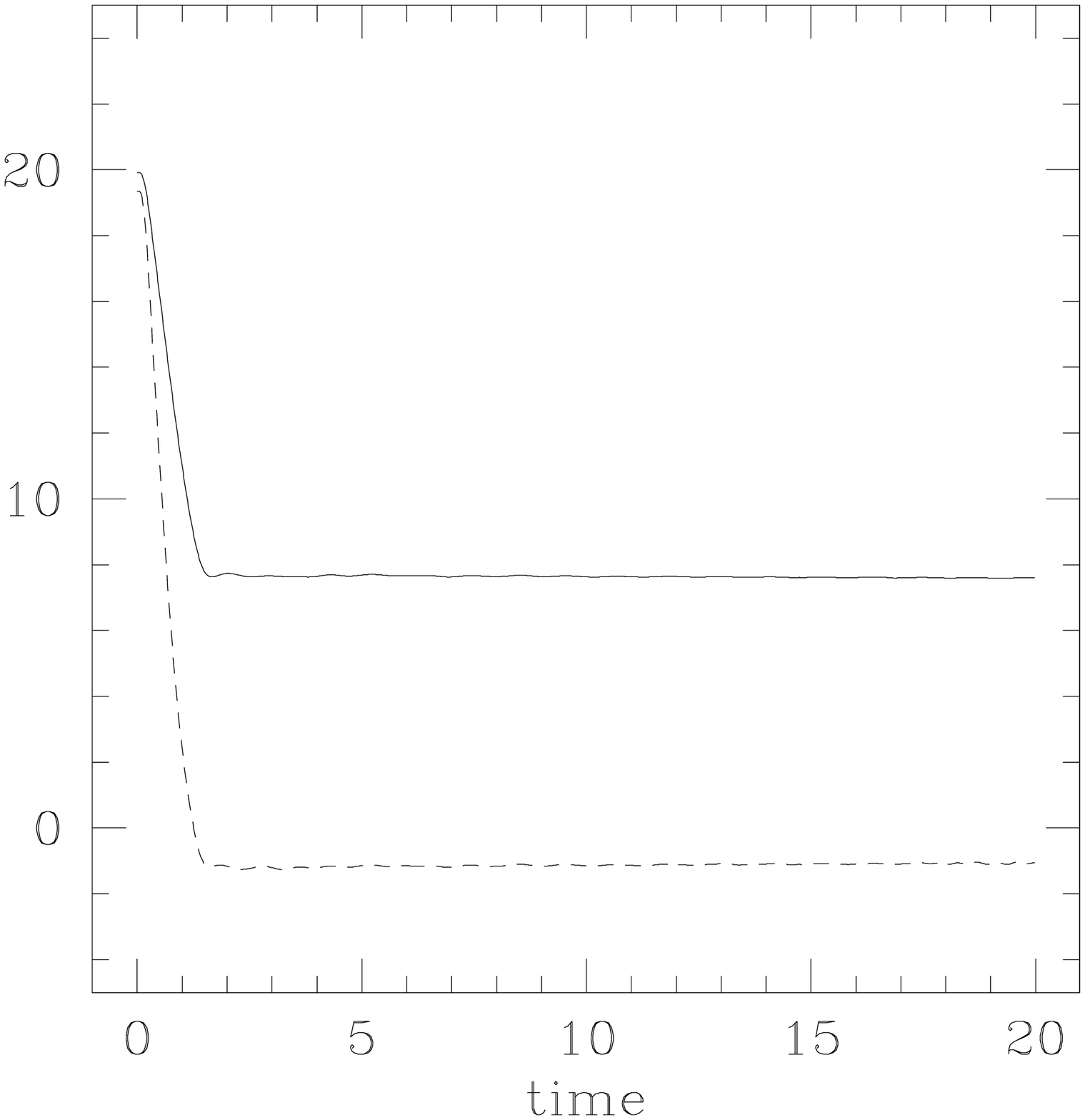,height=6.5cm}
  }}
  \vspace{0.25in}
  \caption{On the left: four steps in the process of a soliton mediated conformational transition
  for a toy protein. A NLS soliton, moving in the upward direction along the backbone,
  interacts with the curvature field and cause transition to the minimal energy conformation.
  In this example, the initial conformation is locally metastable everywhere along the polymer,
  and this state is characterized by one constant curvature value. The minimal energy state is
  characterized by another value. The transition is therefore from one helical structure to 
  another.
  On the right: the collective momentum (continuous line) and energy (dashed line) of the
  soliton as a function of time, during the folding process. The results were obtained by
  numerical simulation with the following parameter: $m=0.5,\epsilon=2.0,
  \delta=-0.5,\Lambda=3.0$.}
  \label{init_order}
\end{figure}

For the numerical study of our model, it would be useful to define a collective coordinate and 
momentum for a soliton
\begin{equation}
  Q = {1\over M}\int_{-\infty}^{\infty}dx\,x |a|^2\hspace{1in}
  P = -i\int_{-\infty}^{\infty}dx\,(a^*a_x-a^*_xa),
\end{equation}
where $M=\int_{-\infty}^{\infty}dx\,|a|^2$ is the soliton mass (it is a constant of motion). $Q$
and $P$ are in fact canonical variables which represent the soliton `center of mass' and its 
conjugate momentum. The relation $\dot{Q}=P/M$ holds. In the exact NLS equation, $P$ (and 
therefore $\dot{Q}$) is a constant of motion. Here we use $P$ as an indicator for the soliton 
kinetic energy. Simulations reveal that when $\delta=0$
({\it i.e.\/} symmetric barrier) the soliton slows down, its amplitude decreases, and eventually
it disappears. Next, $\delta$ is decreased so that the $\phi=-1$ conformation
becomes lower in energy. When the interaction is strong enough, and the soliton has enough energy 
(actually, when its initial momentum $P$ is large enough), the soliton transfers enough
energy to the conformation angles to ensure dynamical activation of the folding transition to
the local ground state ($\phi=-1$) in practically every point it passes (see 
fig.~\ref{init_order}). However, the
soliton extracts back some of the conformation energy. After a short period of time it
reaches a steady state in which its average velocity is almost constant (it may have some periodic 
fluctuations), and the energy gained balances the energy lost. A similar result was also obtained
with the SG model.

\begin{figure}
  \centerline{\hbox{
    \psfig{figure=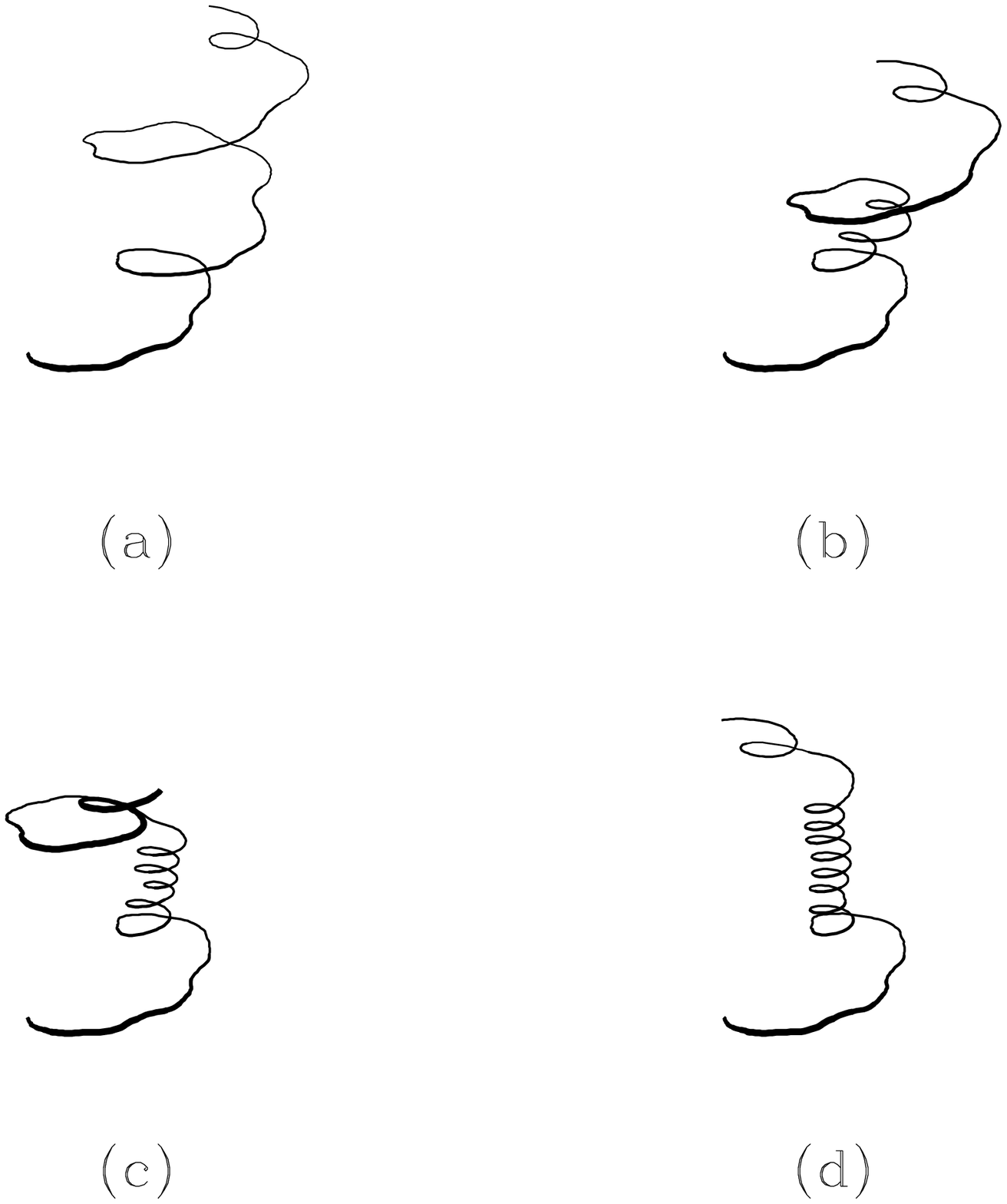,height=6.5cm}
    \hspace{0.2in}
    \psfig{figure=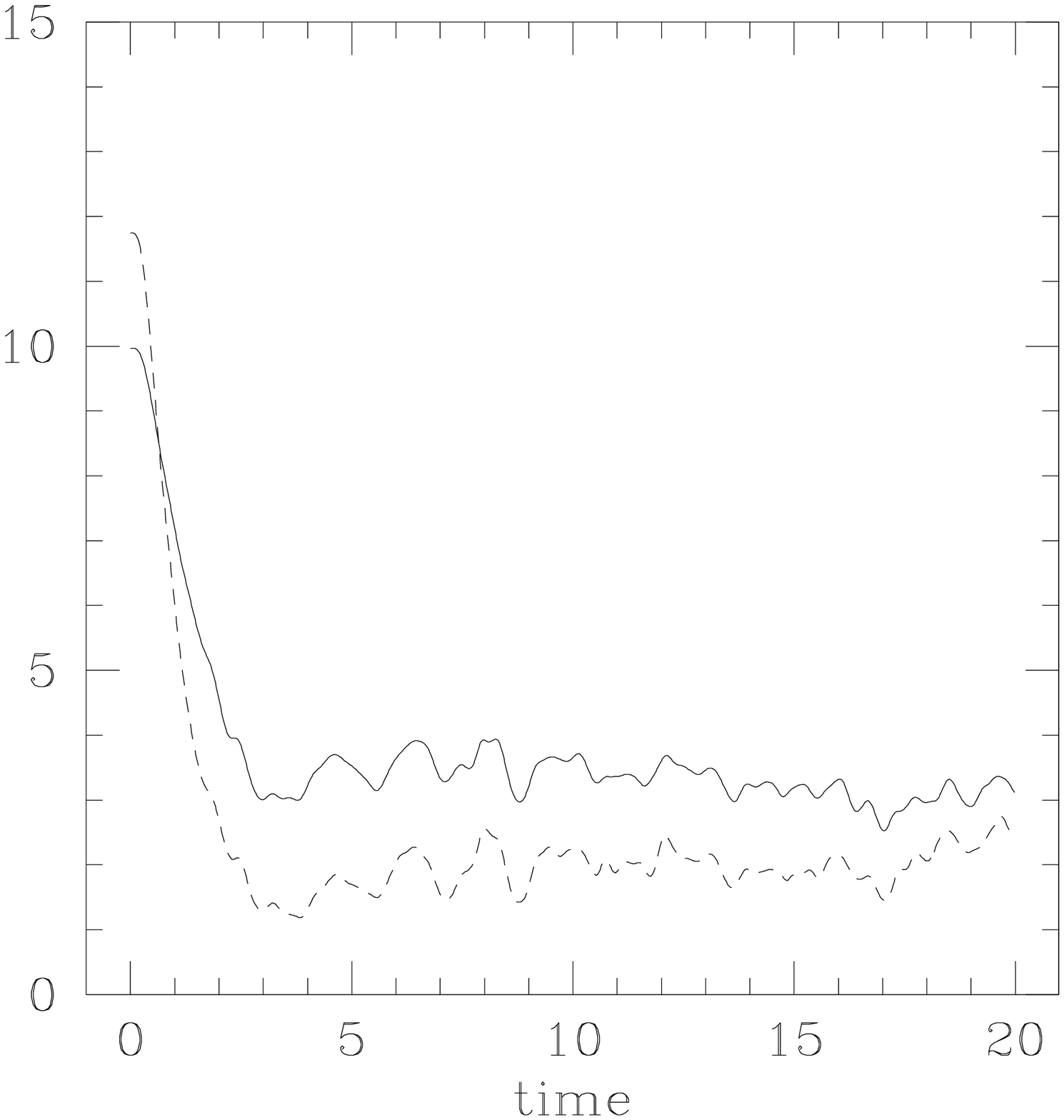,height=6.5cm}
  }}
  \vspace{0.25in}
  \caption{On the left: four steps in the process of soliton mediated folding 
  for a toy protein with an initial random distribution of the stable and the metastable values 
  for the curvature at different points.
  Since in this example the minimal energy curvature is homogeneous, a helical structure is
  mediated. On the right:  the collective momentum (continuous line) and energy (dashed line) as a 
  function of time. The results were obtained in a numerical simulation, with the same parameters as in
  fig.~\protect{\ref{init_order}}.
  }
\label{init_rand}
\end{figure}

In an actual unfolded state local folding angles are supposed to be
distributed somehow between $\alpha$ and $\beta$ conformation. We have checked
the behaviour of a traveling soliton in disordered initial conditions, namely,
when the $\phi$ field is randomly distributed between the two minima. Still,
a steady state velocity is reached, though lower compared with the ordered case, since less
energy is gained as half of the conformation points are already at their ground state. The
folding process is, nevertheless, extremely effective even in this case. The collective momentum
of the soliton during the folding process is shown on the right side of fig.~\ref{init_rand}. 
Four stages of folding are shown on the left side of that figure. 
The field $\phi$ was interpreted as the curvature (after
some rescaling and shifting) of the toy-protein, where the two minima of the potential $V(\phi)$
correspond to two distinct values of curvature. The transformation from a quasi random 
conformation to an ordered helical structure is mediated by a soliton moving in the upward 
direction.

\section{Conclusions}
In this paper we demonstrated that non-topological solitons can participate in a SMCT
process, at least in our much simplified toy-model of proteins, which, nevertheless, captures
some generic properties of the actual molecules. The continuous NLS equation was used as an 
approximate model for Davidov's soliton. We introduced an appropriate interaction (of a 
rather natural form) between the solitons and the conformational degrees of freedom,
which correspond to the $\psi,\phi$ angles of real proteins. The energy that was carried by the
soliton is transfered into the conformation field, and this allows it to overcome energy barriers
and reach the desired ground state. The soliton then receives back some of the energy that was
gained in this process, which, at the ``steady state'', balances its energy loss. This process
can go on as long as enough local folding energy is available. Unlike a stochastic process, in
which all the energy that is gained at a specific local fold transition dissipates, here a much
more efficient mechanism allows a fast and well-controlled conformation transition.

We can now suggest a possible scenario for protein folding. This process may be composed from
few fast SMCT events which cause creation of local structures (the precise nature of which is 
determined, nevertheless, also by global interactions and constraints). 
Every event by itself does not cause an immediate change of the global structure, since large 
amplitude movements of the entire heavy molecule are much slower. The global reorganization of
the molecular conformation should therefore be a slow relaxation of strains which follows the
SMCT. This sets up global conditions for a new SMCT process. Note that this scenario provides
a rather deterministic folding pathway. Soliton creation is probably induced spontaneously by 
thermal excitation, or by an external agent such as `chaperone' enzyme. It might be that solitons
are generated in locations of preferable amino-acids sequences. Their propagation might be 
blocked on different sequences. Therefore, the sequence may not only dictate the final 
conformation, but also the dynamics of the conformational transition.

Conformation changes can probably be mediated by a single SMCT event, followed by slow 
relaxation. We should note that SMCT processes are always exothermic. Conformation changes 
which are associated with increase in energy and entropy, such as thermal protein denaturation,
are not be explained by this model, but do not need complicated mechanism since denaturated 
state is not specific.  The SMCT transition we described is always 
from a metastable to a stable state. In order for that transition
to occur in the reverse direction, external interactions should raise the energy of the initial
state above that of the target state. We shall elaborate on this matter elsewhere.

This research is partly supported by a GIF grant.

\end{document}